\documentclass[%
aip,
jmp,%
amsmath,amssymb,groupaddress,
reprint,
]{revtex4-1}

\usepackage{graphicx}% Include figure files
\usepackage{dcolumn}% Align TABLE columns on decimal point
\usepackage{bm}% bold math
\usepackage{color}

\begin{document}

\title{Does the configurational entropy of polydisperse 
particles exist?}

\author{Misaki Ozawa}

\author{Ludovic Berthier}

\affiliation{Laboratoire Charles Coulomb, 
UMR 5221 CNRS-Universit\'e de Montpellier, Montpellier, France}

\begin{abstract}
Classical particle systems characterized by continuous size polydispersity, 
such as colloidal materials, are not straightforwardly described
using statistical mechanics, since fundamental issues may arise from 
particle distinguishability. Because the mixing 
entropy in such systems is divergent
in the thermodynamic limit we show that the configurational entropy
estimated from standard computational approaches 
to characterize glassy states also diverges. This reasoning would
suggest that polydisperse materials cannot undergo a glass transition, 
in contradiction to experiments. We explain that this argument stems from 
the confusion between configurations 
in phase space and states defined by free energy minima, and propose 
a simple method to compute a finite and physically meaningful 
configurational entropy in continuously polydisperse systems. 
Physically, the proposed approach relies on 
an effective description of the system as 
an $M^*$-component system with a finite $M^*$, for which 
finite mixing and configurational entropies are obtained.
We show how to directly determine $M^*$ 
from computer simulations in a range of 
glass-forming models with different size polydispersities,
characterized by hard and soft interparticle interactions, 
and by additive and non-additive interactions. 
Our approach provides consistent results 
in all cases and demonstrates that the configurational entropy of 
polydisperse system exists, is finite, and can be quantitatively
estimated.
\end{abstract}

\maketitle

\section{Configurational entropy
in polydisperse systems}

\label{sec:Intro}

Colloidal systems play an important role in a wide spectrum of 
soft condensed matter physics, from crystallization kinetics to the 
glass transition phenomenon. An important motivation is that colloids 
can be seen as ``big atoms'', which 
may prove useful in terms of microscopic observations and
particle design~\cite{poon2004colloids}.
These colloidal particles are composed of a 
very large number of atoms, such that each colloid in a given 
experimental batch is a unique object. Therefore, colloidal particles 
are distinguishable classical objects, which rises  
subtle theoretical issues for their statistical mechanical
treatment~\cite{cates2015celebrating}. In particular, 
colloidal particles are often characterized by a continuous distribution 
$f(\sigma)$
of particle diameters, $\sigma$. Thus, in principle, colloids can be 
readily distinguished by their sizes and there are 
no two identical particles in the system.
Polydisperse systems are widely employed in studies of the glass 
transition because polydispersity efficiently prevents crystallization.
Therefore, understanding how polydispersity impacts the 
theoretical treatment of glass formation is an important 
question that constitutes the main theme of this paper.

The statistical mechanics of continuously polydisperse systems has been widely studied in a number of contexts~\cite{salacuse1982polydisperse,briano1984statistical,warren1998combinatorial,sollich1998projected,sollich2001predicting}.
Particle distinguishability may cause in particular 
some subtle issues, such as the well-known Gibbs paradox~\cite{swendsen2008gibbs} and this requires special attention~\cite{cates2015celebrating}.
In that case, particle indistinguishability stemming from a 
quantum mechanical treatment cannot be invoked for colloidal 
particles~\cite{frenkel2014colloidal}. 

Before proceeding to our discussion of disordered materials, 
let us recall what happens in 
the case of ordered materials.
In a recent article, Cates and Manoharan discuss the existence of 
a colloidal crystal made of weakly polydisperse hard 
spheres~\cite{cates2015celebrating}:
{``In the fluid, the spheres can easily swap places whereas in 
the crystal, they cannot.
For indistinguishable particles, the entropy gain
on transforming from liquid to crystal is extensive. 
The additional entropy
cost of localizing distinguishable particles onto un-swappable
lattice sites contains a term $k_{\rm B} \ln(N!)$ 
where $N!$ counts particle permutations. 
This term must be paid to collapse an accessible phase-space volume 
in which distinguishable particles can change places, 
into one where they cannot. This putative entropy cost is supra-extensive 
($k_{\rm B} N \ln N$) and thus for large $N$
outweighs the extensive entropy on formation of the crystal. Thus
the kinetic approach to entropy predicts that colloidal crystals are
thermodynamically impossible. Yet they are observed every day.''}
This paradox is resolved by including the $N!$ distinct crystal 
{\it configurations} generated by the permutation of $N$ distinguishable 
particles in a single crystalline {\it state}. 
Then, this $N!$ multiplicity term cancels exactly the putative entropy 
cost due to particle distinguishability~\cite{cates2015celebrating}.
Physically, this theoretical treatment corresponds to 
describing the polydisperse system as behaving 
effectively as a one-component system and it
highlights the conceptual difference between {\it configurations} in 
phase space volume and {\it states} defined from free energy minima, 
which are  usually constructed from a much 
larger number of configurations. These basic ideas will
reappear throughout our discussion of disordered materials. 

Another well-known issue arising in continuously 
polydisperse systems is the divergence of the mixing entropy, 
$S_{\rm mix}$~\cite{salacuse1982polydisperse,zhang1999optimal,frenkel2014colloidal}. Mathematically, this is because the mixing entropy of a discrete
mixture of particles diverges in the limit of an infinite 
number of components, which is needed to formally represent
a continuous distribution in the thermodynamic limit.  
In several situations, this divergent contribution is 
immaterial, for instance when discussing 
phase equilibria with a fixed particle size distribution 
$f(\sigma)$~\cite{sollich2001predicting} or mixing processes of two 
different continuous polydisperse 
systems~\cite{paillusson2014role}, because the 
absolute value of the entropy is irrelevant and only entropy differences 
between two systems 
matter~\cite{barratold,sollich2001predicting,paillusson2014role,frenkel2014colloidal}.
In the above-mentioned cases, the diverging contributions to 
$S_{\rm mix}$ cancel each other.
 
All these issues are relevant for a proper treatment of the 
glass transition in polydisperse materials, but this has never 
been carefully discussed before.
The configurational entropy $S_{\rm conf}$ is the central quantity to 
describe the thermodynamics of supercooled liquids approaching the 
glass transition~\cite{kauzmann1948nature,adam1965temperature,berthier2011theoretical}.
Conventionally in computer simulations (and also in experiments), 
$S_{\rm conf}$ is defined as $S_{\rm conf}=S_{\rm tot}-S_{\rm vib}$, where $S_{\rm tot}$ is the total entropy of the system and $S_{\rm vib}$ is the entropy of the 
glass (or vibrational) state where only particle vibrations due to thermal 
fluctuations take 
place~\cite{kauzmann1948nature,adam1965temperature,berthier2011theoretical}.
Physically, $S_{\rm conf}$ quantifies the number of available amorphous states 
in glassy systems. These metastable states are understood as long-lived 
free energy minima~\cite{berthier2011theoretical}, even though their mathematical definition 
in finite dimensional glass models remains 
problematic~\cite{biroli2001metastable}.
The mean-field theory of the glass transition is based 
on the existence of an ideal glass transition occurring 
at temperature $T_{\rm K}$ at which $S_{\rm conf}$ 
vanishes~\cite{charbonneau2014fractal}.
Hence, there is a need for a careful investigation of the entropy 
of polydisperse glassy materials.

An extension of the argument developed by Cates and Manoharan 
for the polydisperse crystal \cite{cates2015celebrating} can be made
for the configurational and mixing entropies of a polydisperse glass.     
In that case, the system transforms 
from the fluid where particle 
swaps are easy to the glass where they are not. 
In multi-component glass-formers,
the entropy of the fluid state must contain a mixing entropy contribution
$S_{\rm mix}$, because the particles can exchange their positions~\cite{coluzzi1999thermodynamics,sciortino1999inherent,sastry2000evaluation,angelani2007configurational}.
It is conventionally  assumed that $S_{\rm vib}$ does 
not contain the mixing contribution~\cite{sciortino2005potential}, 
because particles in the glass state mainly vibrate around a fixed
averaged position in a given glass configuration and particle 
exchanges are neglected. Pushing further this argument, 
the divergence of $S_{\rm mix}$ for continuous polydispersity 
provokes the divergence of the entropy of the fluid,
whereas the glass entropy remains finite. Logically, then, 
if the configurational entropy is infinite in the fluid, 
it cannot vanish and the glass transition cannot exist.
This reasoning is of course in direct contradiction 
with experiments and simulations, where ``polydisperse glasses 
are observed every day'', to paraphrase Cates and 
Manoharan.

The goal of this paper is to provide a conceptually-correct 
and computationally-simple way to define and measure 
the configurational entropy in 
continuously polydisperse systems. There exist many separate studies 
of the glassy behavior of colloidal and polydisperse systems both in
experiments~\cite{hunter2012physics,gokhale2016deconstructing} and 
simulations~\cite{santen2000absence,brumer2004numerical,kawasaki2007correlation,hermes2010thermodynamic,zaccarelli2015polydispersity} 
and of the configurational entropy in glass-forming models~\cite{speedy1993entropy,sciortino1999inherent,sastry2001relationship,angelani2007configurational,berthier2014novel,asenjo2014numerical,martiniani2016turning,vink2002configurational,anikeenko2008structural,zhou2016structural,ronceray2016from}.
However, we are not aware of any detailed theoretical or numerical treatment
of the configurational entropy for continuously polydisperse 
systems, for which standard methods are not valid, as we just 
showed. Our paper thus fills this gap. 

Our basic idea is to treat a continuously polydisperse system as 
an effective $M^*$-component system with a {\it finite} number of 
components $M^*$, which can take real values. 
Within this approach, the glass entropy then also contains 
a diverging contribution which cancels the one in 
the total entropy $S_{\rm tot}$. In other words, we estimate 
a finite and physically-relevant contribution to the mixing entropy, 
which we call an effective mixing entropy $S_{\rm mix}^*$. 
As a result, we obtain a finite estimate 
for the configurational entropy $S_{\rm conf}$ as well.
This treatment thus shows that the configurational
entropy of polydisperse systems is always finite, 
and may possibly vanish at a glass transition. 
In addition, we propose a numerical method to 
determine $M^*$ and $S_{\rm mix}^*$, which is 
generically applicable to any glass-forming computer model.

Before closing this introduction, we need to mention
that recently, an alternative method to compute the configurational 
entropy was proposed~\cite{berthier2014novel}, which is
based on the Franz-Parisi construction~\cite{franz1997phase}. In this approach,
the configurational entropy is estimated as a free energy 
difference, and it is measured in monodisperse and polydisperse
systems in the exact same way. Thus, this 
method automatically produces finite values of the configurational entropy 
without suffering from a divergent mixing entropy. 
Results using this method for a continuous polydisperse system
have only very recently been obtained~\cite{berthier2016inpreparation}.
Because this approach
is numerically more demanding than the standard ones discussed 
in this paper, it is useful  
to make the latter applicable to polydisperse systems, such that
both types of methods can be compared. 

The paper is organized as follows.
In Sec.~\ref{sec:theory}, we propose our theoretical idea to compute the 
configurational entropy in continuous polydisperse system.
In Sec.~\ref{sec:numerical}, we describe the 
numerical method to determine $M^*$ and $S_{\rm mix}^*$.
In Sec.~\ref{sec:results}, we validate our approach for several 
simulation models.
In Sec.~\ref{sec:conclusion}, we provide a 
discussion of our results together with some perspectives.

\section{Theoretical considerations: The main problem and
its physical resolution}
\label{sec:theory}

In this section, we present our approach  
to compute the configurational entropy in continuously polydisperse systems.
We first fix the notations and give some definitions.
We then introduce a simple theoretical model of a binary 
mixture illustrating the core 
problem to be faced.
Through the discussion and resolution of the paradoxical
results obtained in this simple model, we draw more general 
conclusions which lead to our simple proposal
to analyze the configurational entropy of generic polydisperse systems.

\subsection{Mixing entropy in polydisperse systems}

\begin{figure}
\includegraphics[width=8.5cm]{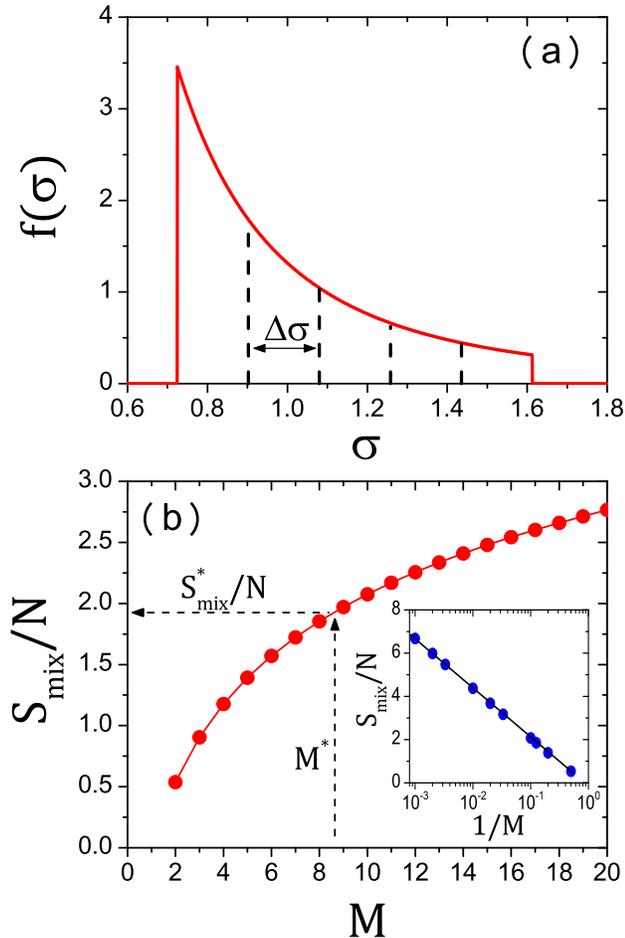}
\caption{
(a) Particle size distribution $f(\sigma) = A\sigma^{-3}$
for $\sigma \in [\sigma_{\rm min}, \sigma_{\rm max}]$, 
$f(\sigma)=0$ outside this range. 
The dashed lines illustrate the decomposition of $f(\sigma)$ 
into the arbitrary example of  
$M=5$ species, with $\Delta \sigma=(\sigma_{\rm max}-\sigma_{\rm min})/M$.
(b) The mixing entropy $S_{\rm mix}$ computed from 
Eqs.~(\ref{eq:S_mix_def}, \ref{eq:X_m}).
The dashed arrows define the value of $S_{\rm mix}^*$
from an effective value $M^*$.
The inset demonstrates the logarithmic 
divergence of $S_{\rm mix}/N$ when $M \to \infty$.}
\label{fig:distribution}
\end{figure}  

We first provide definitions and notations related to
the statistical mechanics and mixing entropy in mixtures of particles.
We consider a continuously polydisperse 
system whose particle size distribution is given by $f(\sigma)$.
To derive the mixing entropy of this system, we decompose $f(\sigma)$ 
into an integer number $M$ of species~\cite{salacuse1982polydisperse,zhang1999optimal,frenkel2014colloidal}, as illustrated in 
Fig.~\ref{fig:distribution}(a) for a specific example.
The continuously polydisperse system is recovered in the limit 
$M\to \infty$.

We consider the canonical ensemble characterized by the number of particles $N$, the volume $V$, and the temperature $T=1/\beta$. 
We set the Boltzmann constant to unity.
The partition function $Z_{\rm tot}$ of the system is
\begin{equation}
Z_{\rm tot} = \frac{1}{ \Pi_{m=1}^{M}N_m ! \Lambda^{dN}} \int_{V} \mathrm{d} {\bf r}^N e^{-\beta U({\bf r}^N)},
\end{equation}
where $\Lambda$, $d$, and $U$ are the de Broglie thermal wavelength, spatial dimension, and potential energy, respectively.
For simplicity, we consider equal masses (hence equal $\Lambda$) irrespective of species throughout this paper.
$N_m$ is the number of particles within species $m$, 
such that $N=\sum_{m=1}^{M} N_m$.
Note that the term, $1/ \Pi_{m=1}^{M}N_m !$ (or more simply $1/ N !$ 
when $m=1$), should be included irrespective of distinguishability of the particles in order to obtain an extensive 
free energy~\cite{warren1998combinatorial,swendsen2008gibbs,frenkel2014colloidal,cates2015celebrating} (see also recent numerical verifications in 
Refs.~\onlinecite{asenjo2014numerical,martiniani2016turning}). 
The total free energy of the system is 
\begin{equation}
\beta F_{\rm tot} = - \ln Z_{\rm tot}.
\end{equation}
Then, the total entropy $S_{\rm tot}$ can be written as 
\begin{equation}
S_{\rm tot}=\beta(E_{\rm tot}-F_{\rm tot}) \sim  \ln Z_{\rm tot} \sim - \ln (\Pi_{m=1}^{M}N_m !),
\end{equation}
where $E_{\rm tot}$ is the total energy of the system.
In this paper, we use the symbol ``$\sim$'' to single out the contribution 
due to the mixing entropy. 
Instead, we use the symbol ``$\simeq$'' to express the 
conventional ``nearly equal'' relation.
By applying Stirling's approximation, $\ln x! \simeq x \ln x - x$ ($x \gg 1$), we get
\begin{equation}
\ln (\Pi_{m=1}^{M}N_m !) \simeq ( N \ln N - N ) + N \sum_{m=1}^{M} X_m \ln X_m,
\label{eq:Stirling}
\end{equation}
where $X_m = N_m / N$ is the concentration of species $m$.
The second term in Eq.~(\ref{eq:Stirling}) comes from the 
effect of having a system with $M$ components, {\it i.e.}, from the 
size polydispersity.
The mixing entropy $S_{\rm mix}^{(M)}$ is defined as 
\begin{equation}
S_{\rm mix}^{(M)} = - N \sum_{m=1}^{M} X_m \ln X_m.
\label{eq:S_mix_def}
\end{equation}
For latter purposes, we note that the mixing entropy 
provides a contribution to the total entropy $S_{\rm tot}$. Using 
the `$\sim$' notation introduced above, we have 
\begin{equation}
S_{\rm tot} \sim S_{\rm mix}^{(M)}.
\label{eq:S_tot_has_S_mix} 
\end{equation}

We can clearly see that in a continuously polydisperse case, 
when we have $M=N$ and hence $X_m=1/N$, the mixing entropy 
$S_{\rm mix}^{(M)}$ diverges in the thermodynamic limit, $N \to \infty$,
\begin{equation}
\frac{S_{\rm mix}^{(M=N)}}{N} = - \sum_{m=1}^{N} \frac{1}{N} \ln \frac{1}{N} = 
\ln N \xrightarrow[N \to \infty]{} \infty.
\end{equation}

Alternatively, if we represent a system characterized by a continuous
size distribution $f(\sigma)$ as an $M$-component 
mixture, then we divide the distribution into $M$ equal intervals, 
$\Delta \sigma=(\sigma_{\rm max}-\sigma_{\rm min})/M$, see 
Fig.~\ref{fig:distribution}(a) for an example. Therefore we have 
\begin{equation}
X_m = \int_{ \sigma_m }^{\sigma_{m+1} } \mathrm{d} \sigma f(\sigma),
\label{eq:X_m}
\end{equation}
where $\sigma_m = \sigma_{\rm min} + (m-1) \Delta \sigma$.
Using Eqs.~(\ref{eq:S_mix_def}) and (\ref{eq:X_m}), we 
get $S_{\rm mix}$ as a function of $M$.
In Fig.~\ref{fig:distribution}(b), 
we show $S_{\rm mix}^{(M)}$ for a specific example of $f(\sigma)$ 
which will be studied numerically below. 
In the inset of Fig.~\ref{fig:distribution}(b), we confirm that 
$S_{\rm mix}/N$ diverges logarithmically in the continuous polydisperse 
limit, $M \to \infty$ (or $\Delta \sigma \to 0$)~\cite{salacuse1982polydisperse,zhang1999optimal,frenkel2014colloidal}, as expected.

Note that in Refs.~\onlinecite{salacuse1982polydisperse,zhang1999optimal,frenkel2014colloidal},
a formal expression of the continuous polydisperse limit is given by
\begin{eqnarray}
\lim_{M \to \infty} \frac{S_{\rm mix}^{(M)}}{N} &=& - \lim_{\substack{M \to \infty \\ \Delta \sigma \to 0}} \sum_{m=1}^{M} f(\sigma_m) \Delta \sigma \ln (f(\sigma_m) \Delta \sigma) \nonumber \\ 
&=&  - \int_{0}^{\infty} \mathrm{d} \sigma f(\sigma) \ln f(\sigma) - \lim_{\Delta \sigma \to 0} \ln \Delta \sigma. 
\label{eq:ill_defined}
\end{eqnarray}
In this expression, the divergent mixing entropy is decomposed into 
a finite term depending on $f(\sigma)$ (the first term) 
and a divergent term (the second term) in Eq.~(\ref{eq:ill_defined}).  
However, the first term is not a well-defined quantity, since it
depends on the argument or labeling of the distribution 
$f$~\cite{zhang1999optimal}, and hence one cannot use the first 
term alone for a physically relevant mixing entropy to 
compute the configurational entropy.

\subsection{A paradox for binary mixtures}

We introduce a simple model of a glass-forming binary mixture 
whose analysis produces 
paradoxical results that illustrate the core of the problem
that we face regarding the mixing entropy of polydisperse particles.
The physics of this system has been discussed qualitatively 
in Ref.~\onlinecite{coluzzi1999thermodynamics}.

We consider a binary mixture ($M=2$) composed of species 1 and 2 in arbitrary dimensions.
The diameters of the particles 
are respectively
$\sigma_{\rm 1}=\sigma$ and $\sigma_{\rm 2}=\sigma(1 + \epsilon)$, 
with $\epsilon > 0$. 
For simplicity, we consider equal concentrations, $X_1 = X_2 = 1/2$.
In this setting, the total entropy $S_{\rm tot}$ always contains the mixing entropy $S_{\rm mix}$ irrespective of the absolute 
value of $\epsilon$, so that the mixing
contribution to the total entropy reads
\begin{equation}
S_{\rm tot} \sim S_{\rm mix} = N \ln 2.
\label{eq:S_tot_paradox}
\end{equation}

In previous studies of binary mixtures with $\epsilon \simeq 1$, this mixing entropy term was included to compute the configurational entropy 
and to detect the ideal glass transition~\cite{speedy1998hard,sciortino1999inherent,coluzzi1999thermodynamics,sastry2000evaluation,angelani2007configurational,ozawa2015equilibrium}.
However, the fact that the mixing entropy does not depend on $\epsilon$ 
is quite surprising. Indeed, 
when $\epsilon \ll 1$, one can expect that the physics of such a 
nearly monodisperse system should not be very different from that of 
purely one-component system obtained by setting $\epsilon=0$. 
In particular, the ideal glass transition temperature $T_{\rm K}$ 
(or density $\phi_{\rm K}$ for hard spheres) 
of this binary system, if it exists, should not 
be very different from that of the purely one-component system. 
More precisely, one would expect that $T_K$ is a continuous function 
of $\epsilon$ near $\epsilon=0$.
However, the configurational entropy $S_{\rm conf}/N$ of the binary system 
is larger than that of the purely one-component system by 
$\ln 2 \simeq 0.7$, due to the mixing entropy contribution 
in Eq.~(\ref{eq:S_tot_paradox}), and this large entropy 
appears discontinuously as soons as $\epsilon > 0$. The theoretical analysis
of this system is therefore physically inconsistent.

One possible solution to the paradox would be to always neglect
the mixing entropy contribution, as done for instance in 
Ref.~\onlinecite{biazzo2009theory}.
However, such treatment would change the location of the 
ideal glass transition significantly.
Let us consider two examples to illustrate this point.
In the Kob-Andersen (KA) model~\cite{kob1995testing}, 
the mixing entropy contributes 
$S_{\rm mix}/N \simeq 0.5$ to the total entropy 
$S_{\rm tot}$~\cite{sciortino1999inherent}.
If one neglects $S_{\rm mix}/N$ from $S_{\rm conf}/N$, 
the estimated $T_{\rm K} \simeq 0.3$~\cite{sciortino1999inherent,sastry2000evaluation,ozawa2015equilibrium} increases to 
$T_{\rm K} \simeq 0.5$, a temperature that is actually 
easily equilibrated in the simulations~\cite{kob1995testing}.
In the binary hard sphere model studied in 
Ref.~\onlinecite{angelani2007configurational}, one has 
$S_{\rm mix}/N = \ln 2 \simeq 0.693$~\cite{angelani2007configurational}.
If one again removes $S_{\rm mix}/N$, the estimated ideal glass transition 
density, $\phi_{\rm K} \simeq 0.62$~\cite{angelani2007configurational}, 
is now reduced to $\phi_{\rm K} \simeq 0.58$, where equilibrium is again 
easily achieved~\cite{angelani2007configurational}.
Assuming that the other contributions to the 
numerical estimate of $S_{\rm conf}$ are well-controlled, then 
it appears impossible to neglect the mixing entropy.

These examples demonstrate that the mixing entropy term 
is actually needed for a proper estimate of the configurational entropy.
But if so, then it must be included in any binary mixture, 
including one where $\epsilon \ll 1$. But in that case the limit
$\epsilon \to 0$ produces a jump in the estimated location of the 
glass transition, which is unphysical. 

In summary, when considering our simple example of a binary mixture
with size ratio $(1+\epsilon)$, taking $\epsilon$ as a continuous 
parameter suggests that previous approaches to determine the configurational
entropy cannot be applied directly, as one needs to decide {\it a priori} 
whether or not to include the mixing entropy in the
configurational entropy. Such decision can still produce 
meaningful results when $\epsilon$ is a constant which is either
large enough or very small. But when intermediate $\epsilon$ values 
are considered, or when the particle size distribution becomes 
more complex, an {\it adhoc} approach becomes impossible.
Importantly for the present work, this approach
breaks down completely when the polydispersity becomes continuous, because
then a continuous spectrum of large and small $\epsilon$ values 
coexists in a single system.

\subsection{Physical resolution of the 
binary mixture paradox}

To resolve the above paradox, we carefully discuss 
the vibrational (or glass) state and its entropy.
We suppose that the equilibrium system is sufficiently deeply 
supercooled that there is a strong separation 
of timescales between vibrations and structural relaxation. This implies
that it makes sense to consider a vibrational state. 

To compute the vibrational entropy $S_{\rm vib}$, one must evaluate the partition 
function of this vibrational state, $Z_{\rm vib}$, by performing a configurational integral within the restricted phase space volume $V_{\rm basin}$ 
explored by the system over the vibrational timescale.
To perform the configurational integral, a representative basin, denoted by $\alpha$, is randomly selected and then the configurational integral is 
evaluated numerically using methods such as the harmonic approximation 
around the local minimum of the potential energy~\cite{sciortino1999inherent,sastry2000evaluation}, or a Frenkel-Ladd~\cite{frenkel1984new}
thermodynamic integration over a system constrained by 
harmonic springs~\cite{coluzzi1999thermodynamics,sastry2000evaluation,angelani2007configurational}.

However, there is a factorial degeneracy of $\alpha$ 
in the phase space volume. In particular, 
these basins can be generated by permutation of the 
particles~\cite{cates2015celebrating}. 
Thus, we have to take into account the number of physically identical 
basins to evaluate $Z_{\rm vib}$.
When $\epsilon \simeq 1$ (corresponding to typical
binary mixtures~\cite{sciortino1999inherent,angelani2007configurational}),
the multiplicity is simply given by the number of configurations
obtained by permuting particles within each species
(1 $\leftrightarrow$ 1 or 2 $\leftrightarrow$ 2), and this corresponds to $N_{\rm 1}! N_{\rm 2}!$.
Instead, when $\epsilon \ll 1$, the permutation of particles 
between unlike species (1 $\leftrightarrow$ 2) should also
be taken into account, because this manipulation also generates configurations belonging to the physically identical basin.

Thus, we need to multiply the configurational integral
by $N!$ instead of $N_{\rm 1}! N_{\rm 2}!$.
As a result, $Z_{\rm vib}$ is different in the two cases, 
$\epsilon \simeq 1$ and $\epsilon \ll 1$, as follows,
\begin{equation}
Z_{\rm vib} = 
\left\{
\begin{aligned}
\quad &\frac{ N_{\rm 1}! N_{\rm 2}!  }{N_{\rm 1}! N_{\rm 2}! \Lambda^{dN}}  
\int_{V_{\rm basin}} \mathrm{d} {\bf r}^N e^{-\beta U({\bf r}^N)}  \quad \left(\epsilon \simeq 1 \right),   \\
\quad &\frac{N!}{N_{\rm 1}! N_{\rm 2}! \Lambda^{dN}} \int_{V_{\rm basin}} \mathrm{d} {\bf r}^N e^{-\beta U({\bf r}^N)}  \quad \left(\epsilon \ll 1 \right).
\end{aligned}
\right.
\end{equation}
Therefore, the vibrational entropy contains a mixing entropy 
contribution whose value depends explicitly on the value of $\epsilon$, 
\begin{equation}
S_{\rm vib} \sim \ln Z_{\rm vib} \sim  
\left\{
\begin{aligned}
\quad & 0   \qquad \left(\epsilon \simeq 1 \right)   \\
\quad & N \ln 2  \qquad \left(\epsilon \ll 1 \right).
\end{aligned}
\right.
\label{eq:S_vib_paradox}
\end{equation}
When $\epsilon \simeq 1$, the vibrational entropy does not contain
a finite mixing entropy, which is consistent with 
previous studies~\cite{sciortino1999inherent,angelani2007configurational}.
However, when $\epsilon \ll 1$, the vibrational entropy contains 
a mixing entropy contribution, which originates from 
the fact that particles with similar (but distinct) 
sizes can be permuted within the vibrational state.
By combining Eqs.~(\ref{eq:S_tot_paradox}, \ref{eq:S_vib_paradox}), 
we obtain $S_{\rm conf}$ as,
\begin{equation}
S_{\rm conf} = S_{\rm tot} - S_{\rm vib} \sim   
\left\{
\begin{aligned}
\quad & N \ln 2   \qquad \left(\epsilon \simeq 1 \right)   \\
\quad & 0  \qquad \left(\epsilon \ll 1 \right).
\end{aligned}
\right.
\end{equation}
In this reasoning, the mixing entropy contribution to $S_{\rm vib}$ 
for $\epsilon \ll 1$ exactly cancels the one in $S_{\rm tot}$
and thus $S_{\rm conf}$ does not contain any mixing entropy, 
which is consistent with the physics of a purely one-component system.
As a result, $T_{\rm K}$ would not change discontinuously 
between $\epsilon \ll 1$ and the purely one-component system
with $\epsilon=0$, thus resolving the paradox.
Note that recently the binary mixture paradox has been resolved also by a first-principle computation~\cite{ikeda2016note}.

What do we learn from this very simple example?
First, a binary system can be described by either an effective 
one-component system or by a truly binary system, depending on 
the value of $\epsilon$.
In practice, these two cases should be distinguished by investigating
the effect of physically swapping the position of particles belonging to 
different species (1 $\leftrightarrow$ 2).
One can equivalently consider that the positions of the particles
are fixed, while that their diameters are swapped, assuming that the mass of each species is identical. 
These two views are of course identical, but swapping diameters is 
more convenient to describe the response of the system 
to such swaps. From such an analysis, one should be 
able to determine a critical value $\epsilon^*$ of $\epsilon$ such 
that mixtures with diameter ratio characterized 
by $\epsilon < \epsilon^*$ should be treated as one-component
systems, whereas for $\epsilon > \epsilon^*$ a binary description 
would be needed. Evaluating $\epsilon^*$ from a first-principle analysis
is a worthwhile goal~\cite{ikeda2016note}. 

To determine which case is appropriate to describe a given binary mixture,
one should determine whether the system remains within the same 
original basin after swapping the particle diameters between 
different species. If it does, then an effective one-component 
description is appropriate, otherwise a binary description is needed. 
In both cases, $S_{\rm vib}$ and $S_{\rm conf}$ contain a mixing entropy
contribution, but its precise value depends on which 
effective description is selected by the above procedure. 
Therefore, the general conclusion is that one must always include 
$S_{\rm mix}$ into $S_{\rm vib}$, but its value must be 
numerically determined by a careful analysis of the 
robustness of the basin to particle diameter swaps. 

\subsection{General strategy: Mapping to an effective 
$M^*$-component system}

Our main idea is that a continuously polydisperse system 
should be regarded as an effective $M^*$-component system
with $M^* < \infty$. Such a simple idea has been 
used before for instance to study phase equilibria (see 
Ref.~\onlinecite{sollich2001predicting} and references therein) and 
prediction of the equation of 
state~\cite{ogarko2012equation,ogarko2013prediction}.
In the following, we show how to apply this idea 
to obtain a meaningful definition of the configurational entropy.

We consider a generic polydisperse system in a sufficiently 
deeply supercooled liquid regime, where the vibrational entropy
can be well-defined. The particle size distribution is characterized by 
$M$ species, where $M=\infty$ if the polydispersity is continuous.

The physical idea that particles 
with diameter ratio smaller than a threshold 
(which we called $(1+ \epsilon^*)$ in the above example
of a binary mixture) should be treated 
as identical suggests that a continuous particle 
size distribution could be ``coarse-grained'' into finite bins 
to be effectively treated as an $M^*$-component mixture. 
The value of $M^*$ should be determined by requesting that 
(i) particle diameter swaps within a single effective species 
($m$ $\leftrightarrow$ $m$) leave the basin unaffected
(ii) particle diameter swaps between different species 
($m$ $\leftrightarrow$ $n$ with $m \neq n$) drive the 
system out of the original basin.

Once the value of $M^*$ is estimated, then we 
evaluate the partition function within the vibrational state, $Z_{\rm vib}$, 
by taking into account the multiplicity of 
physically identical basins, $\Pi_{m=1}^{M^*}N_m !$, such that
\begin{equation}
Z_{\rm vib} = \frac{\Pi_{m=1}^{M^*}N_m ! }{ \Pi_{m=1}^{M}N_m ! 
\Lambda^{dN}} \int_{V_{\rm basin}} \mathrm{d} {\bf r}^N e^{-\beta U({\bf r}^N)}.
\end{equation}
Note that we use $M$ and $M^*$ in this expression, which respectively
represent the actual and effective numbers of components in the polydisperse
system.  

Using Eqs.~(\ref{eq:Stirling}, \ref{eq:S_mix_def}), 
the vibrational entropy $S_{\rm vib}$ is then given by 
\begin{equation}
S_{\rm vib} \sim \ln Z_{\rm vib} \sim S_{\rm mix}^{(M)} - S_{\rm mix}^{(M^*)}.
\label{eq:S_vib_has_S_mix}
\end{equation}
Note that $S_{\rm vib}$ contains two mixing entropy terms, the 
possibly divergent term, $S_{\rm mix}^{(M)}$, and the finite 
contribution $S_{\rm mix}^{(M^*)}$.
 
Finally, by combining Eq.~(\ref{eq:S_vib_has_S_mix}) with the total entropy $S_{\rm tot}$ in Eq.~(\ref{eq:S_tot_has_S_mix}), we obtain the configurational entropy $S_{\rm conf}$, 
\begin{equation}
S_{\rm conf} = S_{\rm tot} - S_{\rm vib} \sim S_{\rm mix}^{(M^*)}.
\label{eq:final}
\end{equation}
The remarkable result is of course that 
the term $S_{\rm mix}^{(M)}$ which diverges for continuous polydispersity 
has disappeared from this final result, in which 
only the finite value $S_{\rm mix}^{(M^*)}$ remains.
We call effective mixing entropy $S_{\rm mix}^* = S_{\rm mix}^{(M^*)}$
this physically relevant contribution to the configurational entropy.

It is clear from Eq.~(\ref{eq:final}) that 
the argument against the existence of an ideal glass 
transition for continuous polydispersity is now avoided because 
$S_{\rm conf}$ is always finite and should contain a finite 
contribution stemming from an effective mixing entropy. 
Also, the above reasoning demonstrates that the configurational entropy should 
be interpreted or quantified as the (logarithm of the) number of {\it states}, 
and not of {\it configurations}. 
In the inherent structure approach~\cite{stillinger1982hidden} one would count the (infinite) 
number of configurations (or their corresponding inherent structures) 
generated by the mixing effect, which would thus result in a diverging  
configurational entropy~\cite{stillinger1999exponential}.
Instead here, each vibrational `state' accounts for an infinite
number of configurations (and thus of inherent structures).  

In addition, our theoretical strategy provides useful guidelines for 
estimating numerically the effective mixing entropy $S_{\rm mix}^*$ 
for any generic polydisperse glass-forming material. 
In our approach, $M^*$ and thus $S_{\rm mix}^*$ are not 
arbitrary choices, but they should be determined 
by the system itself through the response of the basins to diameter swaps. 
We now describe how to implement these ideas in practice. 

\section{Numerical implementation}
\label{sec:numerical}

\subsection{General algorithm for 
the numerical determination of $M^*$}
\label{subsec:numerical_detection}

As explained in Sec.~\ref{sec:theory}, 
$M^*$ should be such that 
(i) particle diameter swaps within a single effective species 
($m$ $\leftrightarrow$ $m$) leave the basin unaffected
(ii) particle diameter swaps between different species 
($m$ $\leftrightarrow$ $n$ with $m \neq n$) drive the 
system out of the original basin.

\begin{figure}
\includegraphics[width=8.5cm]{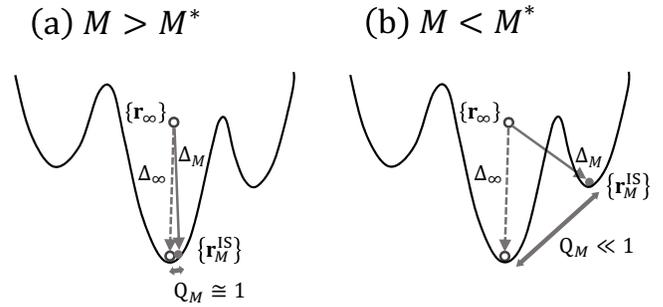}
\caption{Sketch of the potential energy landscape showing the 
initial equilibrium configuration $\{ {\bf r}_{\infty} \}$,
and the inherent structure $\{ {\bf r}_{M}^{\rm IS} \}$ 
obtained after swapping the diameters within each $M$ species. 
(a) When $M>M^*$, the system stays in the same basin so that
$Q_{M} \simeq 1$ and $\Delta_M \simeq \Delta_{\infty}$. 
(b) When $M<M^*$, the system moves to another basin so that 
$Q_{M} \ll 1$ and $\Delta_M \gg \Delta_{\infty}$. 
$M^*$ is defined from the crossover between 
situations (a) and (b).}
\label{fig:schematic_fig}
\end{figure}  

To determine $M^*$, we prepare equilibrium configurations of the original
continuously polydisperse system characterized by a distribution $f(\sigma)$. 
We denote by $\{ {\bf r}_\infty \}$ the original configuration, see 
Fig.~\ref{fig:schematic_fig}.
In order to characterize whether the system stays physically in the 
same basin or 
moves to another basin as a result of the swaps, we perform a quench 
of the system into the inherent structure (IS)~\cite{stillinger1982hidden}. 
The IS corresponding to the initial configuration
is denoted as $\{ {\bf r}_\infty^{\rm IS} \}$.

We then decompose $f(\sigma)$ into $M$ species,
from $M=1$ where the system is treated as an one-species 
system, to $M=\infty$ which describes the original system.
In practice we stop at the large 
value $M=100$.
For a given $M$ value, we 
systematically perform diameter swaps within each species
($m$ $\leftrightarrow$ $m$, with $m=1, \cdots, M$). 
More precisely, for a given configuration, we randomly pick a 
pair of particles within the same species and exchange their 
diameters. 
We repeat such diameter swap $N$ times so that most of the particles 
in the configuration experience the swap.
Once the swaps have been performed, we quench the obtained configuration 
to its IS, which we denote by $\{ {\bf r}_M^{\rm IS} \}$.
Our goal is to monitor whether the system lands in a different
basin (for $M<M^*$) or not (for $M>M^*$), as sketched in
Fig.~\ref{fig:schematic_fig}.
To this end we monitor the following three quantities.

\begin{itemize}

\item The IS energy $e_{\rm IS}$~\cite{stillinger1982hidden}.
When the system moves to another basin, $e_{\rm IS}$ should 
change from the value it has in the original basin.

\item The overlap $Q_{M}$ between the IS configurations 
$\{ {\bf r}_{\infty}^{\rm IS} \}$ and $\{ {\bf r}_{M}^{\rm IS} \}$: 
\begin{equation}
Q_{M} = \frac{1}{N} \sum_{i,j}^{N} \theta( a- | {\bf r}_{M i}^{\rm IS} 
- {\bf r}_{\infty j}^{\rm IS} | ),
\label{eq:overlap}
\end{equation} 
where $\theta(x)$ is the Heaviside step function, and $a$ is a 
microscopic threshold lengthscale. By definition, $Q_{\infty}=1$, 
whereas we expect that $Q_M$ decreases from $1$ with decreasing $M$. 

\item 
The mean-squared displacement 
between the original equilibrium configuration 
$\{ {\bf r}_{\infty} \}$ and the swapped IS $\{ {\bf r}_M^{\rm IS} \}$:
\begin{equation}
\Delta_{M}=\frac{1}{N} \sum_{i=1}^{N} | {\bf r}_{M i}^{\rm IS} 
- {\bf r}_{\infty i} |^2,
\label{eq:MSD}
\end{equation}
so that $\Delta_\infty$ corresponds approximately to the plateau value of the 
mean-squared displacement reached in the physical 
dynamics of the original system.

\end{itemize}

If the system falls into a different basin, one expects 
that $Q_{M}$ will strongly decay from 1 and that $\Delta_M$ will 
increase significantly from $\Delta_\infty$, as shown in  
Fig.~\ref{fig:schematic_fig}. 
When $M>M^*$, the system remains in the same basin after the swaps 
and the quench, showing that $Q_{\rm M} \simeq 1$ and $\Delta_M \simeq \Delta_{\infty}$. 
On the other hand, when $M<M^*$, the original basin is destroyed, and the system moves to another basin, showing that $Q_{\rm M} \ll 1$ and $\Delta_M \gg \Delta_{\infty}$. We expect that $M^*$ can be defined as a crossover value 
between these two extreme cases.

More precisely, we perform the following algorithm to determine $M^*$. \\

{\bf General Algorithm: Swap and quench protocol}
\begin{itemize}{\tt
\item[1)] Define $M$ species by dividing the distribution $f(\sigma)$ into 
$M$ finite bins.
\item[2)] Prepare an equilibrium configuration of the original continuous polydisperse system, $\{ {\bf r}_{\infty} \}$.
\item[3)] Randomly pick up a pair of particles within the same 
bin, and exchange their diameters. Do $N$ such exchanges.
\item[4)] Quench the system and obtain $\{ {\bf r}_{M}^{\rm IS} \}$.
\item[5)] Repeat 1) - 4) for a large number of configurations for 
a wide range of $M$ values, between $M=1$ and $M=100$.
\item[6)] Plot $e_{\rm IS}$, $Q_{M}$ and $\Delta_{M}$ as a function of $M$ and determine $M^*$ as a crossover value.}
\end{itemize}

This ``swap and quench'' protocol is a natural way to determine $M^*$ which follows from the theoretical idea described in Sec.~\ref{sec:theory}.
In this protocol, the distribution of the particle diameters 
$f(\sigma)$ is unaffected after defining $M$ and after the diameter
swaps.

We have considered the following alternative scheme
where one literally performs 
a discretization of the diameters into $M$ species and monitor the effect 
of this coarse-graining on the basin.
However, we found that this coarse-graining process may affect the 
thermodynamic state point of the system for soft potentials.
Therefore, we use this method only for hard sphere systems so that 
the volume fraction (and therefore 
the thermodynamic state point) does not change.
For completeness we also present this alternative 
``coarse-graining and quench'' protocol.
Most of this procedure is as in the above ``swap and quench'' protocol
except for item 3) which is replaced by:
\begin{itemize}
{\tt \item[3')] Discretize
the diameters of the original system into $M$ discrete 
species keeping the global volume fraction constant.}
\end{itemize}
We have checked that the two protocols 
give essentially the same results for hard spheres, as shown below. We mention 
the coarse-graining procedure as a way to trigger the imagination
of future researchers towards the implementation of 
numerical procedures alternative to ours.

{It is important to realize that the discretization 
of the particle size distribution in step 1) of the algorithm
is not uniquely defined. For the particular case under study, 
it is physically meaningful to choose  equal intervals 
$\Delta \sigma = (\sigma_{\rm max}-\sigma_{\rm min})/M$, such that each 
discretized species occupies the same fraction of the total volume,
but one can imagine that non-uniform intervals could be better
choices for more complicated particle size distributions.
We have not addressed this point in detail, but it 
would deserve further study.}

\subsection{Details of our models and simulations}

We perform molecular simulations for several 
pair potentials in three dimensions~\cite{allen1989computer}, 
studying a total of five different numerical models.

We use a continuous size polydispersity, 
where the particle diameter $\sigma$ of each particle is randomly drawn from 
the following particle size distribution: 
$f(\sigma) = A\sigma^{-3}$, for $\sigma \in [ \sigma_{\rm min}, 
\sigma_{\rm max} ]$, 
where $A$ is a normalization constant. 
We define the size polydispersity 
as $\delta=\sqrt{\overline{\sigma^2} - \overline{\sigma}^2}/\overline{\sigma}$, where $\overline{\cdots}=\int \mathrm{d} \sigma f(\sigma) (\cdots)$.
We mainly use $\delta = 0.23$, 
choosing $\sigma_{\rm min} / \sigma_{\rm max} = 0.4492$ when studying
different pair potentials, as shown 
in Fig.~\ref{fig:distribution}. To investigate the effect of 
$\delta$ we also study the case  
$\delta = 0.176$ choosing $\sigma_{\rm min} / \sigma_{\rm max} = 0.543$
for the hard sphere potential.
We use $\overline{\sigma}$ as the unit length. 
We simulate systems composed of $N$ particles in a cubic 
cell of volume $V$ with periodic boundary conditions~\cite{allen1989computer}. 
The canonical ensemble is used throughout this paper.

We use the following pairwise potential for the soft sphere models,
\begin{eqnarray}
v_{ij}(r) &=& v_0 \left( \frac{\sigma_{ij}}{r} \right)^n + c_0 + c_1 \left( \frac{r}{\sigma_{ij}} \right)^2 + c_2 \left( \frac{r}{\sigma_{ij}} \right)^4, \label{eq:soft_v} \\
\sigma_{ij} &=& \frac{(\sigma_i + \sigma_j)}{2} (1-\eta |\sigma_i - \sigma_j|), \label{eq:non_additive}
\end{eqnarray}
where $v_0$ is the unit of energy, and $\eta$  
quantifies the degree of non-additivity of the particle 
diameters. Non-additivity is introduced 
for convenience, as it prevents more efficiently crystallization
and thus it enhances glass-forming ability of the numerical models.
The constants, $c_0$, $c_1$ and $c_2$, are chosen so that the first and 
second derivatives of $v_{ij}(r)$ become zero at the cut-off 
$r_{\rm cut}=1.25 \sigma_{ij}$.
We employ the additive and non-additive soft sphere models 
using the parameters $n=18$, $\eta=0$ (additive) 
and $n=12$, $\eta=0.2$ (non-additive), respectively.
We set the number density $\rho=N/V=1.0$ with $N=1500$ for the soft 
sphere models.

We also use hard sphere models in 
three dimensions, where the 
pair interaction is zero for non-overlapping particles, and infinite otherwise. 
We employ both additive ($\eta=0$) and non-additive ($\eta=0.2$) 
systems of hard spheres, using Eq.~(\ref{eq:non_additive}).
Also, we analyze an additive hard sphere system with $\delta=0.176$.
We use $N=1000$ for the additive hard sphere systems and $N=300$ 
for the non-additive systems.
We have also performed simulations using $N=300$ and $8000$ for additive hard 
spheres with $\delta=0.23$ 
and find that there is no significant finite size effects in our results.
The hard sphere simulations are presented as a function of the 
reduced pressure $Z=P/(\rho k_B T)$, where $P$ is the measured 
pressure, and $k_B T$ is set to unity. 

The initial equilibrium configurations for the protocols described 
in Sec.~\ref{subsec:numerical_detection} need to be deeply supercooled 
either at very low temperatures (for soft sphere potentials) 
or high pressures (for hard spheres). 
To this end, we employ a swap Monte-Carlo (MC) method for all models.
This approach is a very efficient thermalization algorithm which enables 
us to easily reach sufficiently supercooled 
equilibrium regimes. 
The details of these efficient simulations are provided in 
Ref.~\onlinecite{ninarello2016inpreparation} for the soft sphere models 
and Ref.~\onlinecite{berthier2016equilibrium} for the hard sphere models.

Finally to perform the quench of the system into its 
inherent structure, we use a conjugate gradient 
method~\cite{nocedal2006numerical} for the soft sphere systems.
To realize the quench for the hard sphere systems, we perform 
non-equilibrium compressions up to the jamming volume fraction 
$\phi_J$~\cite{stillinger1985inherent,chaudhuri2010jamming,ozawa2012jamming}.
Specifically we employ the jamming algorithm described in 
Ref.~\onlinecite{xu2005random,desmond2009random} to perform these 
nonequilibrium compressions.

\section{Numerical validation of the method}
\label{sec:results}

\subsection{Practical determination of $M^*$}

\begin{figure*}
\includegraphics[width=17cm]{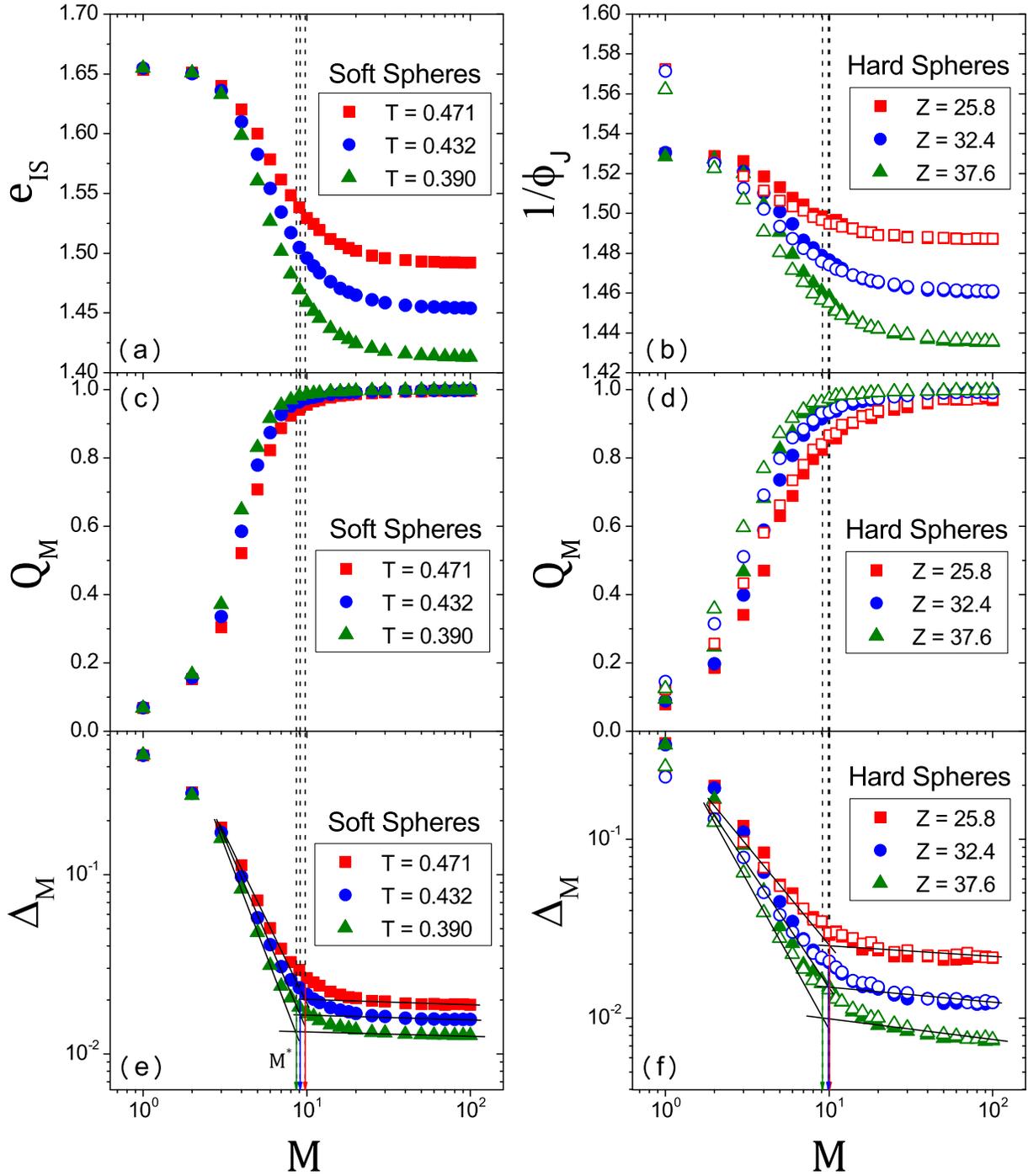}
\caption{
(a, b) Inherent structure energy $e_{\rm IS}$ and volume $1/\phi_{\rm J}$.
(c, d) Overlap $Q_{M}$.
(e, f) Mean-squared displacements $\Delta_M$.
The results are for additive soft spheres (left panels) 
and additive hard spheres (right panels) with $\delta=0.23$.
The vertical dashed lines represents the $M^*$ values 
determined by the intersection of the two power law regimes 
in the mean-squared displacements in (e, f). 
In addition to the ``swap and quench'' protocol with closed symbols, 
results for the ``coarse-graining and quench''
protocol are also shown for hard spheres using open symbols.}
\label{fig:IS_MSD_overlap}
\end{figure*}  

We now present our numerical results. 
In Fig.~\ref{fig:IS_MSD_overlap}, we show 
the evolution of the inherent structure energy 
$e_{\rm IS}$ (or the volume at jamming $1/\phi_{\rm J}$ for hard spheres), 
of the overlap $Q_{M}$, and of  
the mean-squared displacement $\Delta_M$ as a function of $M$
for both a soft sphere model and a hard sphere model. All the 
other models we have studied behave similarly. 
To compute $Q_{M}$, we set $a=0.23$ for all systems.
This value was used earlier to characterize metastable states and 
the phase transitions among them~\cite{BJ15,berthier2016inpreparation}.
   
The results of the additive soft sphere model with $\delta=0.23$ are presented in the left panels of Fig.~\ref{fig:IS_MSD_overlap}.
At large $M$, we observe nearly constant values of 
$e_{\rm IS}(M) \simeq e_{\rm IS}(M \to \infty)$, $Q_{M} \simeq 1$, and $\Delta_M \simeq \Delta_{\infty}$, which means that the swap of the diameters 
within each $M$ species marginally affects the system.
Thus,  after the diameter swaps, 
the system essentially remains in the original basin.
However, with decreasing $M$, all observables start to deviate 
significantly from their large-$M$ limits. 
As $M$ decreases, we find that $e_{\rm IS}$ increases, $Q_{M} \to 0$, 
and $\Delta_M$ increases rapidly. 
These observations indicate that at smaller $M$, the effect of the swap is 
so strong that the original basin is destroyed, and the system moves to 
another basin.

It is clear from these figures that a clear crossover 
occurs between large and small $M$ behaviors, which 
we wish to use to determine $M^*$ quantitatively. 
However, because $M^*$ describes a smooth crossover, it cannot
be defined unambiguously. From a careful analysis 
of all these observables for all models, we conclude that $\Delta_M$ is 
the best quantity to determine $M^*$, because it appears as the most
sensitive measure of the change of basin. As shown 
in Fig.~\ref{fig:IS_MSD_overlap}(e), 
we determine $M^*$ as the intersection of two power law regimes 
obeyed at large and small values. 
For the soft sphere model in Fig.~\ref{fig:IS_MSD_overlap}(e), 
we obtain $M^* \simeq 8.5-10$ with a weak temperature 
dependence. These values of $M^*$ are also reported in the 
figures for $e_{\rm IS}$ and $Q_{M}$ by the vertical dashed lines. It is clear that
they identify also very well the crossover behavior 
observed in these two observables, even though their crossover 
behavior is less sharply defined.

In the right panels of Fig.~\ref{fig:IS_MSD_overlap}, 
we show the results for the additive hard sphere system with $\delta=0.23$.
We observe a behavior which is qualitatively similar 
to the soft sphere model, and we determine that 
$M^* \simeq 9-10$ using again the 
$M$-dependence of $\Delta_M$ in Fig.~\ref{fig:IS_MSD_overlap}(f).

In addition, for hard spheres we can also compare these results 
obtained with the ``swap and quench'' protocol, 
to the ``coarse graining and quench'' protocol.
The results for the latter procedure are shown 
with open symbols in Fig.~\ref{fig:IS_MSD_overlap}.
The excellent agreement between the two sets of results is evident
and this second protocol thus yields the same values of 
$M^*$. 

\subsection{Results for $M^*$ and $S_{\rm mix}^*$}

We have repeated the measurements shown in 
Fig.~\ref{fig:IS_MSD_overlap} for the five different numerical
models considered in this work. For each model, we obtain 
$M^*$ for a broad range of deeply supercooled state points. 
Finally, using the simple procedure shown in 
Fig.~\ref{fig:distribution}(b), we convert these $M^*$-values into 
an estimate of the effective mixing entropy $S_{\rm mix}^*$, which we could then use to 
determine the configurational entropy of these continuously 
polydisperse systems.

\begin{figure*}
\includegraphics[width=17cm]{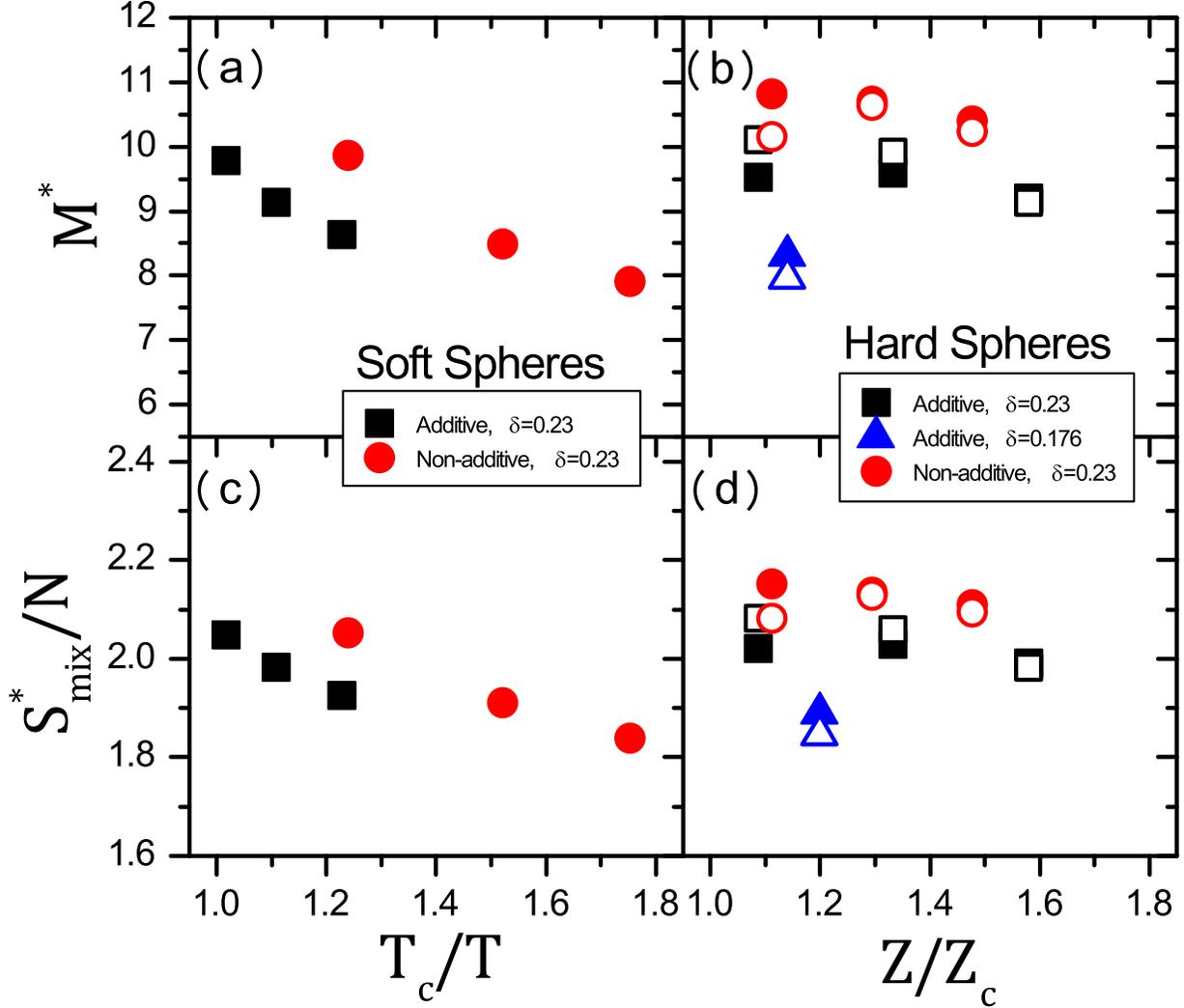}
\caption{Evolution of $M^*$ for two soft sphere systems 
(a) and three hard sphere systems (b).
(c, d) show the corresponding effective mixing entropy
$S_{\rm mix}^*$. The horizontal axes are normalized by the 
mode-coupling transition points, $T_c$ and $Z_c$ to facilitate 
the comparison between the different models.
In addition to the ``swap and quench'' protocol (closed symbols), 
results for the ``coarse-graining and quench''
protocol (open symbols) are also shown for hard spheres.}
\label{fig:M_and_smix}
\end{figure*}  

We compile all our results for $M^*$ and $S_{\rm mix}^*$ 
as a function of $1/T$ (for soft potentials) and $Z$ (for hard spheres)
in Fig.~\ref{fig:M_and_smix}.
The temperature and pressure scales are normalized by 
the mode-coupling transition 
points, $T_c$ and $Z_c$, in order to facilitate the comparison 
between different glass-formers. 
As usual, $T_c$ and $Z_c$ are determined by a power-law fit of the relaxation 
times~\cite{ninarello2016inpreparation,berthier2016equilibrium}.
Notice that all studied state points correspond to fluid states 
thermalized well beyond the mode-coupling crossover. 

We find that $M^*$ and hence $S_{\rm mix}^*$ slightly decrease 
with decreasing $T$ or increasing $Z$, and this effect 
seems more pronounced for the soft sphere models. Physically, 
this reflects the fact that the basin becomes more robust 
against particle exchanges at lower $T$ or higher $Z$.
This variation is a direct confirmation that $M^*$ represents 
a non-trivial characterization of the glassy states, which needs 
to be numerically determined independently for each state point, 
at least in principle.

In practice however, it is a very good approximation to consider 
that the effective mixing entropy $S_{\rm mix}^*$ of hard spheres 
is essentially constant in the deeply supercooled fluid. 
In addition, in the hard sphere systems the two distinct protocols
introduced above give quantitatively consistent values, especially at 
large $Z$, which again confirms the robustness of 
our numerical approach.

Remarkably, the value of $M^*$ for the additive system 
with $\delta=0.176$ is about $20\%$ smaller than that of the system 
with $\delta=0.23$, which may be rationalized by the 
corresponding relative reduction of the 
polydispersity (of about $\approx 23~\%$). Thus, our results
appear in agreement with the physical expectation that 
$(M^*-1)$ should be proportional to the polydispersity $\delta$.
We would need more data at different polydispersities to 
confirm this result quantitatively. However, a rough 
extrapolation of these $M^*$ values towards smaller 
polydispersity values, suggests that $M^* \simeq 4-5$ for $\delta=0.10$ 
and $M^* \simeq 2-3$ for $\delta \simeq 0.05$, whereas smaller 
polydispersities should essentially be treated as mono-component 
systems. We can also translate these findings into a critical 
value $(1 + \epsilon^*) \simeq 1.1$ for the size ratio below which 
two particles should be treated as having the same size.

Finally, we note that for the relatively polydisperse systems that 
we study, the effective mixing entropies are of order $S_{\rm mix}^*/N \simeq 2.0$.
Given that the configurational entropy is typically of order 
$S_{\rm conf}/N \simeq 1.0$ near the onset of slow dynamics~\cite{sciortino1999inherent,angelani2007configurational}, we conclude 
that an accurate determination of the mixing entropy is indeed crucial 
to properly estimate the configurational entropy. 

\section{Discussion and perspectives}
\label{sec:conclusion}

In this paper, we discussed the problems of the conceptual definition 
and of the practical measurement of the configurational entropy $S_{\rm conf}$ in 
glass-forming models characterized by a polydisperse size 
distribution, inspired by the physics of colloidal materials.
We first noticed that standard approaches to estimate the configurational
entropy fail in the case of a continuous 
polydispersity, providing either an infinite or an incorrect 
estimate of $S_{\rm conf}$ depending on an arbitrary choice made
regarding the mixing entropy contribution. 
We then proposed a simple method to 
compute a finite and physically-meaningful configurational 
entropy $S_{\rm conf}$ in continuously polydisperse systems 
which relies on a treatment of the original system 
as an effective $M^*$-component system, where $M^*$ is finite
and has to be measured directly at each state point.
Finally, we performed numerical simulations of five different 
glass-forming models and showed that our approach provides meaningful
results in all cases. 

The key idea leading to a finite configurational entropy
for polydisperse systems is the distinction between 
{\it configurations} (or their corresponding inherent structures) in the phase space volume 
and {\it states} defined as free energy minima. While 
the configurational entropy should be interpreted or quantified as the 
(logarithm of the) number of states, the inherent structure 
approaches instead quantify the (logarithm of the) number of inherent structures associated to configurations. 
While this may represent a correct approximation 
in some specific instances, the confusion between the two concepts
has important consequences for polydisperse mixtures, as the 
number of configurations (and hence of inherent structures) diverges, while the number of states does not.
This conceptual distinction has been discussed at length
in the literature~\cite{biroli2000inherent,berthier2011theoretical}, 
as the confusion between states and configurations is 
 at the root of several arguments suggesting the impossibility of 
an ideal glass 
transition~\cite{stillinger1988supercooled,donev2007configurational,eckmann2008ergodicity}. Therefore, the case of 
polydisperse materials is one more instance where this confusion
may lead to paradoxical results. 

We mentioned in the introduction the recently-proposed 
alternative method to compute the configurational entropy, based 
on the Franz-Parisi construction, which relies on estimating the
free energy difference between the fluid and glass 
states~\cite{franz1997phase,berthier2014novel}. This method 
does not suffer from a possible divergence in the mixing 
entropy, because it does not involve estimating the entropy 
of the fluid and glass states separately. 

In fact, the conceptual difference between the standard approach 
($S_{\rm conf}=S_{\rm tot}-S_{\rm vib}$) 
and the Franz-Parisi construction is easy to grasp
from their practical implementations. Both approaches
rely on performing thermodynamic measurements on systems 
constrained to evolve `near' a given reference configuration. When 
using the Frenkel-Ladd method in the standard approach~\cite{coluzzi1999thermodynamics,sastry2000evaluation,angelani2007configurational}, the constraint
is a set of harmonic springs connecting {\it each particle} to reside close to 
its position in the reference configuration. Instead, in the Franz-Parisi
approach, one constrains the {\it collective density profile} to 
reside close to the one of the reference configuration. For 
the particular case of polydisperse systems, the collective
nature of this constraint allows particle permutations in the glass state
which are then automatically taken into account with their correct
equilibrium weight. 
{Note that the standard definition of the overlap used 
in this approach does not take into account the particle diameters.
One could imagine generalizing this definition to prevent 
particle permutations between particles with very different diameters, 
which would 
result in a different estimate of the configurational entropy, suggesting
that a more careful study of the effect of size polydispersity 
within the Franz-Parisi approach would be very valuable too.}
In order to use the standard approach
we have shown that it is necessary to correct
for the (incorrect) `self' nature of the constraint by estimating separately 
an effective mixing entropy contribution, and we have proposed 
a simple method to do so.

Because both methods provide distinct conceptual estimates
of the configurational entropy, there is no reason 
why they should yield identical absolute values of $S_{\rm conf}$ in the 
supercooled liquid regime. 
In the presence of a genuine ideal glass transition, however, both methods should yield consistent
results for its location.
In Ref.~\onlinecite{berthier2016inpreparation}, we have computed the 
volume fraction dependence of the 
absolute value of $S_{\rm conf}$ for the additive hard sphere system 
with $\delta=0.23$ studied in the present work, 
using both the present approach and the 
Franz-Parisi construction. The results show that the 
agreement between the two methods is good~\cite{berthier2016inpreparation}.

Note that in experiment, the configurational entropy is obtained by 
integration of the heat capacity difference between the fluid and crystal 
states~\cite{richert1998dynamics}, where it is assumed 
that the vibrational 
(or glass) entropy is well approximated by the entropy of the crystal~\cite{yamamuro1998calorimetric,angell2002specific}.
The reference point which sets the absolute value of the 
configurational entropy is given by the entropy of fusion.
Therefore, traditional experimental measurements of the configurational entropy 
do not suffer from the problem of the mixing entropy, because 
the experimental procedure does not require estimating the 
absolute entropy of any state point.
This experimental procedure could be applied to systems with 
continuous polydispersity, because the experimental definition of the 
configurational entropy simply requires a thermodynamic integration
along an equilibrium path.
Of course, 
the drawback of this procedure is that it is unclear whether 
the experimental protocol truly reflects the theoretical definition 
of a configurational entropy related to the number of free energy minima.
Therefore, for theoretical purposes we do not wish to follow the experimental
procedure to solve the issue raised by polydispersity. Similarly, 
the experimental path cannot be followed in computer simulations where
the properties of the equilibrium crystal are usually unknown, 
and thermodynamic integration across the nonequilibrium glass transition 
would yield incorrect results. Therefore, it is numerically 
necessary to carefully estimate the mixing entropy.

{In this paper, we defined $M$ discretized species by dividing the distribution $f(\sigma)$ into equal intervals $\Delta \sigma = (\sigma_{\rm max}-\sigma_{\rm min})/M$, such that each species occupies the same fraction of the total volume for the specific 
distribution shown in Fig.~\ref{fig:distribution}(a).
However, one can expect that non-uniform intervals $\Delta \sigma_m$, could be applied to general distributions $f(\sigma)$.
Whereas non-uniform intervals might change the value of $M^*$ and functional form of $S_{\rm mix}^{(M)}$, we would expect that the resulting $S_{\rm mix}^*$ does not depend on the binning process as long as physically reasonable 
discretizations producing relatively small $M^*$ values are applied.
Finding such reasonable non-uniform intervals for general 
distributions $f(\sigma)$ is an important topic for future 
studies.}

Our numerical protocol, which transforms a continuously 
polydisperse system into an $M^*$-component system, 
is also useful for more general theoretical investigations, since 
finite mixtures are theoretically more tractable than continuously 
polydisperse systems~\cite{coluzzi1999thermodynamics,biazzo2009theory,weysser2010structural}.
For example, theoretical computations of the thermodynamics of 
the glass phase in (finite and discrete) multi-components mixtures~\cite{coluzzi1999thermodynamics,biazzo2009theory}
could more easily be applied once $M^*$ is numerically determined.
In the same vein, in Ref.~\onlinecite{weysser2010structural}, 
the multi-components 
extension of the mode coupling theory is compared with simulations of a 
continuous polydisperse system with $\delta=5.8\%$.
The theory seems to describe well simulation results regarding glassy dynamical 
behaviors when $M \geq 3$, which is in good agreement with the 
estimate obtained in our work assuming that 
$(M^*-1)$ is proportional to $\delta$.
Our work in fact paves the way for an analytical treatment of 
the glass thermodynamics of continuously 
polydisperse particle systems. Another interesting theoretical 
goal would be the analytical derivation of the critical size ratio 
$(1 + \epsilon^*)$ separating the regime where particles should be treated 
as identical or as distinct, as this would allow a fully analytical
treatment of polydisperse materials. 
Recently, a starting point for this direction has been developed~\cite{ikeda2016note}.
When applied to hard spheres, 
predictions could then be made regarding the influence of the 
polydispersity on the jamming transition, which is another
topic of interest.

\begin{acknowledgments}

We thank P. Charbonneau, D. Coslovich, A. Ikeda, 
H. Ikeda, K. Miyazaki, A. Ninarello, P. Sollich, G. Tarjus, and F. Zamponi for helpful comments.
We specially thank A. Ninarello for providing us with
very low temperature configurations for the soft sphere models.
The research leading to these results has
received funding from the European Research Council
under the European Union’s Seventh Framework
Programme (FP7/2007-2013)/ERC Grant Agreement
No. 306845. This work was supported by a grant from the Simons 
Foundation (\# 454933, Ludovic Berthier)

\end{acknowledgments}

\bibliography{smix}

\end{document}